\documentclass[aps,prl,twocolumn,floatfix]{revtex4}
\usepackage{graphicx,units,xspace,amsmath}

\graphicspath{{figures/}}

\newcommand{\micron}{\ensuremath{\unit{\mu m}}\xspace}

\begin{document}

\title{Sorting by Periodic Potential Energy Landscapes: Optical Fractionation}

\author{Kosta Ladavac}
\author{Karen Kasza}
\author{David G. Grier}

\affiliation{Dept.\ of Physics, James Franck Institute and
Institute for Biophysical Dynamics\\
The University of Chicago, Chicago, IL 60637}

\date{\today}

\begin{abstract}
  Viscously damped objects driven through a periodically modulated potential
  energy landscape can become
  kinetically locked in to commensurate directions through the
  landscape, and thus can be deflected away from the driving
  direction.  
  We demonstrate that the threshold for an object to become
  kinetically locked in to an array
  can depend exponentially on its size.  
  When implemented with an array of holographic optical tweezers,
  this provides
  the basis for a continuous and continuously optimized sorting 
  technique for mesoscopic objects called ``optical fractionation''.
\end{abstract}

\pacs{87.80.Cc, 82.70.Dd, 05.60.Cd}
\maketitle

\begin{figure}[t]
  \centering
  \includegraphics[width=\columnwidth]{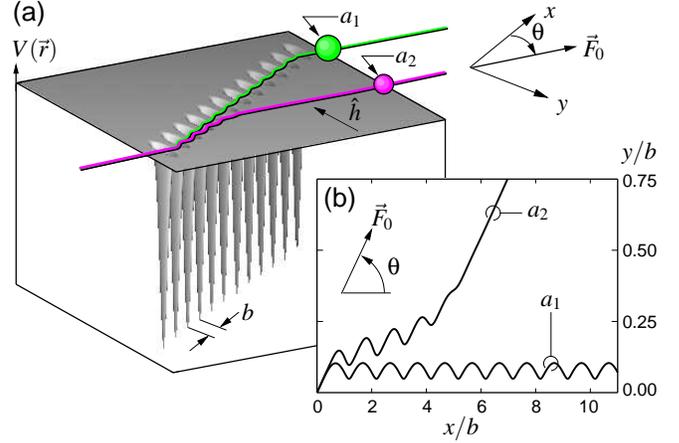}
  \caption{Principle of optical fractionation.  (a) Different types of particles
    are driven by external force $\vec F_0$ through an array of optical traps
    inclined at angle $\theta$ with respect to $\hat F_0$.  Strongly interacting
    particles ($a_1$) become kinetically locked-in to the array and
    deflected, while the others ($a_2$) are not.
    (b) Trajectories for large ($a_1 = 0.79~\micron$) and small
    ($a_2 = 0.5~\micron$) spheres calculated with Eq.~(\ref{eq:potential})
    for experimental conditions described in the text.
  }
  \label{fig:schematic}
\end{figure}
Many natural and technologically important
processes involve classical transport of small 
objects through modulated potential energy landscapes.
While the generic behavior of modulated transport is well understood in
one dimension \cite{risken89}, fundamental questions remain for
higher dimensions.
Colloidal particles flowing through arrays of optical tweezers
\cite{ashkin86,hot}
provide a uniquely accessible experimental archetype for this class of problems.
Experiments on transport through square arrays
have revealed a Devil's staircase hierarchy of
kinetically locked-in states as a function of orientation \cite{korda02b}.
Within each locked-in state, particles select
commensurate paths through the array independent of the driving
direction.
The ability to selectively deflect one fraction out of a flowing
mixture was predicted \cite{korda02b}
to be useful for sorting and purifying mesoscopic samples.
This Letter describes a practical implementation of this process,
which we term optical fractionation.
Examining the kinematics of optical fractionation further
reveals that the underlying lock-in transition can be 
\emph{exponentially} sensitive to size.

\begin{figure*}[t!]
  \centering
  \includegraphics[width=0.75\textwidth]{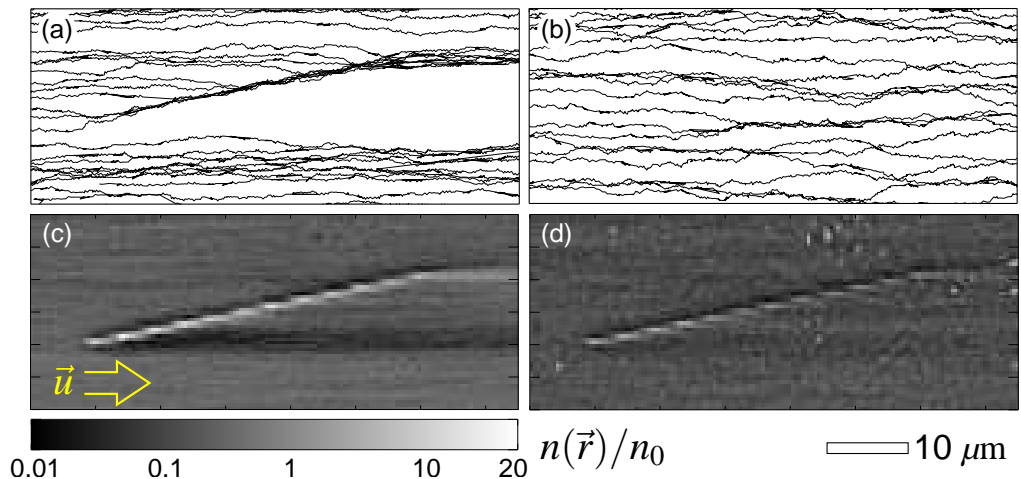}
  \caption{Optical fractionation of bidisperse silica spheres.
  (a) Representative trajectories for $a_1 = 0.79~\micron$ at 1/60~\unit{sec}
  intervals.  (b) Trajectories for $a_2 = 0.50~\micron$ obtained simultaneously.
  (c) Time-averaged areal density $n_1(\vec r)$ for $a_1$ relative to the mean,
  $n_0$.  Data compiled from 30,000 trajectories over 4 hours.  
  (d) Simultaneously acquired data for 0.50~\micron radius spheres compiled from 45,000 trajectories.
  The color bar indicates $n_i/n_0$ for both data sets, and the
  scale bar denotes 10~\micron for all four panels.}
  \label{fig:bigsmall}
\end{figure*}

Optical fractionation exploits a competition between
optical gradient forces exerted by an array
of discrete optical tweezers \cite{ashkin86}
and an externally applied force, as shown in
Fig.~\ref{fig:schematic}.
A driven particle's trajectory can be deflected enough
by an encounter with one optical tweezer to pass into the domain
of the next, and so on down the line.
Such a trajectory is said to be kinetically locked-in to the array.
Optical fractionation's selectivity emerges because objects
with different sizes, shapes or compositions can experience
substantially different potential energy landscapes in the
same optical intensity distribution, and the periodicity of
the lattice emphasizes these differences.
More strongly driven or weakly trapped objects
escape from the array and flow away in the driving direction.
The two resulting fractions can be
collected in separate microfluidic channels downstream.

To demonstrate optical fractionation in practice, we studied the transport
of water-borne colloidal particles flowing past
a linear array of optical tweezers projected with the holographic optical
tweezer technique \cite{hot}.
The colloidal suspension was confined to a 
$4~\unit{mm} \times 0.7~\unit{mm} \times 40~\micron$ glass channel formed by bonding
the edges of two glass cover slips.
Capillary forces at the channel's inlet were used to create a
flow of about $60~\unit{\micron/s}$ along the midplane.
This flow carried a
mixture of monodisperse silica spheres
of radius $a_1 = 0.79~\micron$ (Duke Scientific Lot No.~24169)
and $a_2 = 0.5~\micron$ (Duke Scientific Lot No.~19057), both of which can be tracked
to within 30~\unit{nm} in the plane at 1/60~\unit{sec} intervals 
using digital video microscopy \cite{crocker96}.
The two sizes can be distinguished reliably,
and typical trajectories appear in Fig.~\ref{fig:bigsmall}.

Colloidal silica spheres are roughly twice as dense as water and
settle into a monolayer just above the lower glass wall of the channel, 
with the smaller spheres floating about 1~\micron higher. 
Given the Poisseuille flow profile in the channel, the larger spheres travel
somewhat slower, with a mean speed of $u_1 = 13 \pm 2~\unit{\micron/s}$,
compared with the smaller spheres' $u_2 = 17 \pm 9~\unit{\micron/s}$.
The spheres experience a viscous drag, $\vec F_j = \gamma_j \vec u_j$, characterized by a
drag coefficient, $\gamma_j$, that depends both on their
radii, $a_j$, and on proximity to surfaces
\cite{dufresne01}.

Twelve discrete optical tweezers
were arranged in a line  with center-to-center spacing $b = 3.6 \pm 0.1~\micron$
oriented at $\theta = 12.0^\circ \pm 0.5^\circ$ with respect
to the flow.
Each trap was powered by $1.7 \pm 0.8~\unit{mW}$ of laser light at 532~\unit{nm},
which slightly exceeded the empirically determined
lock-in threshold for the larger spheres, given $\theta$ and $b$.

The trajectories in Fig.~\ref{fig:bigsmall}(a) and (b) demonstrate
that the larger spheres are indeed systematically deflected by the array
of traps, while the smaller spheres are not.
Consequently, the array creates a shadow in the distribution of large spheres
into which the small spheres can flow.
Because the purification of small spheres and concentration of large results
from lateral deflection, this optical fractionation
process can proceed continuously, in contrast to most competing techniques
\cite{wilson00}.

\begin{figure}[t!]
  \centering
  \includegraphics[width=\columnwidth]{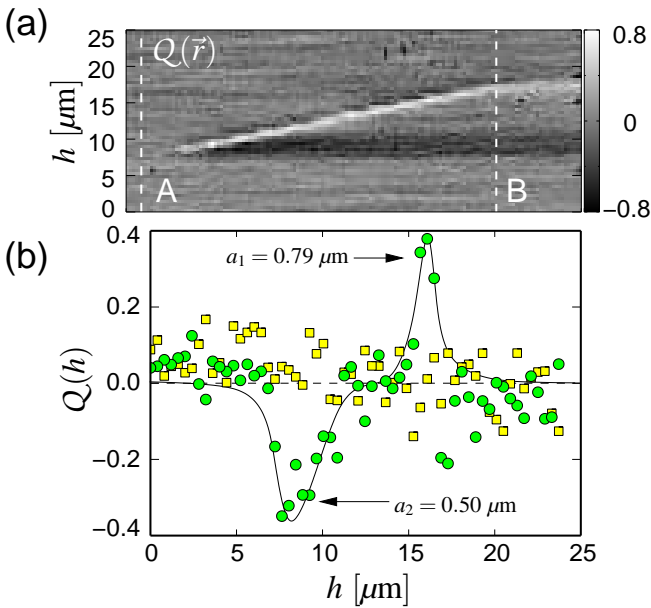}
  \caption{(a) Spatially resolved quality of separation ${\cal Q}(\vec r)$
    obtained with a single line of optical traps.  The cross-section transverse
    to the flow direction along line A is plotted as squares in (b) and
    provides a baseline profile for the
    suspension's composition before optical fractionation.  The cross-section along B,
    plotted as circles in (b) shows the influence of the trap array.  The curve in (b)
    is a guide to the eye.
  }
  \label{fig:qualfig}
\end{figure}

These qualitative observations can be made more
compelling by considering the statistics of tens of thousands of trajectories
compiled in Fig.~\ref{fig:bigsmall}(c) and (d).
Here, we plot the two populations' time-averaged areal densities $n_j(\vec r)$ normalized by
their means.
The separation's quality is assessed in Fig.~\ref{fig:qualfig} with
${\cal Q}(\vec r) \equiv [n_1(\vec r) - n_2(\vec r)]/[n_1(\vec r) + n_2(\vec r)]$,
which reaches unity in regions containing only
large spheres, and minus one in regions with only small spheres.
A transverse section taken along line $A$ in Fig.~\ref{fig:qualfig}(a)
and plotted as squares in Fig.~\ref{fig:qualfig}(b)
reveals a thoroughly mixed sample with ${\cal Q}(h) = 0$ approaching the
traps.
A similar section along line $B$ downstream of the array
demonstrates roughly 40 percent purification
of both large and small spheres.
Much of the background can be attributed to the traps' nonuniform
intensities \cite{korda02b}, with large particles
tending to escape from the weakest traps.
In denser suspensions, this escape rate is increased by collisions.
Both processes can be mitigated by projecting multiple lines of traps
\cite{korda02b}.

Optical fractionation's ability to distinguish objects arises 
as a general and previously unappreciated
feature of transport through periodically structured environments.
Analyzing such transport not only provides insights
into optimizing practical sorting, therefore, but also sheds new light
on a range of analogous processes.

The potential energy landscape presented by an optical trap array
is a convolution of the traps'
intensity profile $I(\vec r)$ with an object's optical form factor $f(\vec r)$: 
$V(\vec r) = - I(\vec r) \circ f(\vec r)$.
The total applied force then is
\begin{equation}
  \label{eq:force}
  \vec F = \vec \nabla [I(\vec r) \circ f(\vec r)] + \vec F_0,
\end{equation}
where $\vec F_0$ is the driving force.
Colloidal particles' motions are overdamped, so that the resulting
velocity is $\vec v = \vec F / \gamma$.
The associated trajectory has a component $v_x$ along the row of traps
and another $v_y$ perpendicular.
Although overdamped transport through a periodic potential
is well understood for one-dimensional systems \cite{risken89},
few analytic results are available for the inclined line,
and fewer still incorporate thermal or quenched randomness.
Consequently, we focus on the kinematic limit in which
both driving and trapping forces exceed random thermal forces
so that trajectories may be treated deterministically.
We then estimate the threshold for an
object to escape from an array of optical traps, and thereby
establish the selectivity of optical fractionation.

Any trajectory entrained by the traps, such as the example in Fig.~\ref{fig:schematic}(b),
is characterized by turning points where $v_y = 0$.
Conversely, any trajectory without such turning points must be unbounded.
This establishes as the maximum possible locked-in
deflection angle
\begin{equation}
  \label{eq:condition}
  \sin \theta_m \le \max\left\{ 
    \partial_y \left[- I(\vec r) \circ f(\vec r)\right] \right\} / F_0.
\end{equation}

At a given $\theta$, objects are either deflected or not
with a selectivity set by the dependence of $\sin \theta_m$
on material properties.
To estimate $\theta_m$, we model
the array as a periodically modulated
line of light with intensity $I_0$:
\begin{equation}
  \label{eq:line}
  I(\vec r) = I_0 \, A(y) \,
  \sum_{n = 0}^\infty \alpha_n \cos (n k x),
\end{equation}
where $k = 2 \pi / b$, 
the dimensionless transverse distribution $A(y$) is
symmetrically peaked around $A(0) = 1$, 
and the coefficients $\alpha_n$ account for the tweezers' detailed
structure with $\sum_n \alpha_n = 1$.

Convolving first along the $x$ direction by applying
the Fourier convolution theorem to each term in the sum,
and then noting that
$\cos nkx \le 1$ yields
\begin{equation}
  \sin \theta_m \le \frac{V_0}{F_0} \, \max \left\{- \partial_y \left[A(y) \circ
      \sum_{n = 0}^\infty 
      \alpha_n
       \tilde f(nka,y) \right] \right\}
\end{equation}
where $V_0 \propto I_0$ is the potential wells' depth and
$\tilde f$ is the form factor's Fourier transform along the $x$ direction.
The array's periodicity thus selects a discrete set of wavenumbers
from the continuous $\tilde f$ whose dependence on $a$ endows
optical fractionation with exceptional size selectivity.
This is most clearly demonstrated if
$\tilde f$ can be factored into inline and transverse components,
$\tilde f(nka,y) = \tilde f_x(nka) f_y(y)$.
In this case,
\begin{equation}
  \label{eq:limit}
  \sin\theta_m \le q(a) \,
  \sum_{n = 0}^\infty \alpha_n \tilde f_x(nka),
\end{equation}
with $q(a) = \kappa(a) V_0/F_0$ and
$\kappa(a) = \max\left\{ - \partial_y [A(y) \circ f_y(y)] \right\}$.
Equivalent results can be obtained
when $\tilde f$ is not separable, but require
a lengthier term-by-term analysis.

We turn our attention first to the transverse contribution.
If a particle is comparable in size to the optical tweezers' width,
$w_0$, then $\kappa(a)$ depends no more strongly on size
than $1/a$.  For example, if $A$ and $f_y$ are Gaussians of
widths $w_0$ and $a$, respectively, then
$\kappa(a) \propto 1 / \sqrt{w_0^2 + a^2}$.
Similarly, the potential depth $V_0$ and driving force $F_0$
scale as simple powers of $a$
in the absence of nonlinear effects such as Mie resonances \cite{bohren83}.
Consequently, the prefactor $q(a)$ describes 
no more than
algebraic sensitivity to size and material properties.
Comparable selectivity is offered by other
techniques such as gel electrophoresis and field flow fractionation \cite{wilson00}, 
and would be obtained with an unmodulated line of light ($\alpha_0 = 1$).

The in-line contribution is more interesting.
Because a particle's form factor vanishes outside the
interval $x \in [-a,a]$, its Fourier transform depends very
strongly on wavenumber.
For example, a uniform dielectric
cube aligned with the array has a separable form factor, with
$f_x(x) = 1$ for $x \in [-a,a]$.
Its Fourier transform,
\begin{equation}
  \tilde f_x(nka) = \frac{\sin nka}{nka},
\end{equation}
is bounded by the leading-order cumulant expansion
\begin{equation}
  \label{eq:exponential}
  \tilde f_x(nka) = \exp\left(- \frac{n^2 k^2 a^2}{6} \right)
\end{equation}
for $ka < \pi$.
The equivalent result for a sphere \cite{bohren83} with
$f(\vec r) = \sqrt{1 - r^2/a^2}$ for $r \in [-a,a]$
is $\tilde f_x(nka) \approx (\pi/2) \exp\left(- n^2 k^2 a^2/6 \right)$,
with $f_y(y) \approx \exp( -n^2 k^2 y^2/6)$.
Any smooth, bounded, positive-definite $f(\vec r)$ on $x \in [-a,a]$ 
similarly would surpass exponential sensitivity
for $ka \gtrsim 1$.
Substituting Eq.~(\ref{eq:exponential}) into Eq.~(\ref{eq:limit})
therefore establishes
optical fractionation's exponential size sensitivity
for $1 \lesssim ka < \pi$.

Comparable sensitivity to control parameters is observed in analogous transitions between
sub-harmonic steps in drive charge density waves \cite{brown84} and between kinetically
locked-in states in two-dimensional optical trap arrays \cite{korda02b}.
Similar results also can be obtained for arrays of potential barriers, suggesting that
arrays of optical tweezers also should be effective for sorting absorbing and reflecting
particles that are repelled by laser light.
This analysis also carries over to filtration by arrays of micromachined posts
\cite{duke97}, which 
therefore should be able to resolve objects substantially smaller than the inter-post
separation under the right circumstances.

Both $\tilde f(nka,y)$ and the
coefficients $\alpha_n$
fall off rapidly with index $n$.
Consequently, the sum in Eq.~(\ref{eq:limit})
is dominated by the first term, $n = 1$.
This improves the approximations used in deriving Eqs.~(\ref{eq:limit})
and (\ref{eq:exponential}) and suggests that the result
may be considered an estimate for $\sin \theta_m$ rather
than simply a bound.

To demonstrate this, we apply this analysis to our present experimental data,
modeling the individual optical traps as Gaussian potential wells
\begin{equation}
  \label{eq:potential}
  V(\vec r) = V_0(a) \sum_{j = 1}^N \exp\left( - \frac{(\vec r - j b \hat x)^2}{2 \sigma^2(a)} \right),
\end{equation}
with $\sigma^2(a) \approx w_0^2 + a^2$ \cite{sigma}.
In this model,
\begin{equation}
  \label{eq:model}
  \sin \theta_m \lesssim q(a) \, \exp \left( - \frac{b^2}{8 \sigma^2} \right),
\end{equation}
where $q(a) = (2/\sqrt{e}) \, V_0 / (\sigma F_0)$.
The weakest trap's occupancy, $n_j$, is inversely
proportional to the particles' minimum speed, $v_\text{min}$,
as they pass through.
Consequently, we can estimate the relative trap strength from the
data in Fig.~\ref{fig:bigsmall} as
$q(a) = 2(1 - v_\text{min}/u)$.
Similarly, the separation between the depleted region ahead of the traps
and the position of maximum occupancy
is $2 \sigma(a)$.
From the histograms in Fig.~\ref{fig:bigsmall}(c) and (d), 
we obtain $q(a) = 1.6 \pm 0.1$ and $0.9 \pm 0.2$, and
$\sigma(a) =  0.85 \pm 0.07~\micron$ and $0.58 \pm 0.07~\micron$
for the large and small spheres, respectively \cite{sigma}.
These results suggest $\theta_m = 14^\circ \pm 1^\circ > \theta $ for the
large spheres and $\theta_m  = 3^\circ \pm 2^\circ \ll \theta$ for the small, 
which is consistent
with the observation that only the large spheres are systematically deflected.

The threshold $\sin \theta_m$ depends only linearly on $V_0$ and $F_0$.
Thus, imperfections in practical trap arrays
and fluctuations in the driving force
can be largely compensated for
by the substantially stronger dependence on particle size.
Indeed, Figs.~\ref{fig:bigsmall} and \ref{fig:qualfig}
demonstrate robust size separation despite
more than 20 percent variation in
flow velocity over the course of the experiment.

Equation~(\ref{eq:model}) also offers insights into applying
optical fractionation to nanometer-scale objects.
Stokes drag
scales linearly with $a$ and the optical trapping potential
for Rayleigh particles scales
as $a^3$, so that $q(a) \propto a^2$.
Sorting proteins or nanoclusters, therefore, will require enhancing
$V_0$ by four orders of magnitude.
This might be accomplished by
increasing the light's intensity, reducing its wavelength \cite{neto00}, 
and exploiting resonances.
Even then,
only algebraic size sensitivity should be expected for
objects with $a \ll \lambda$ because
$ka \ll 1$ in this limit.

We have focused on effects due to
deflection transverse to the optical
axis.  
Multi-dimensional separations could take advantage of Bessel beams' 
ability to exert controlled
axial forces \cite{arlt01} to distribute objects 
both transverse to and along the optical axis.

In summary, we have demonstrated optical fractionation for a model
system of bidisperse colloidal spheres.
This approach lends itself to continuous, rather than batch-mode fractionation,
with continuous tuning and dynamic optimization over the entire size range
accessible to optical trapping, \emph{i.e.} nanometers to micrometers.
The abrupt transition from free flow to kinetically
locked-in transport should offer exponential size selectivity
for objects larger than roughly 100~\unit{nm}.
Separation on the basis of 
other characteristics also can be optimized, although
exponential sensitivity should not be expected in general.
Our analysis focuses on the kinematic limit, $F_0 b > V_0 > k_BT$,
which is both tractable and appropriate for weakly-trapped micrometer-scale
colloid.
Stronger trapping would require a more detailed treatment of thermally assisted
hopping \cite{risken89}.  A substantially more sophisticated analysis also would be required
for higher-dimensional arrays.
Similarly abrupt transitions should occur with variation of driving or trapping strength
for vortex creep through
patterned type-II superconductors \cite{reichhardt99}, 
electron transport through two-dimensional
electron gases \cite{wiersig01}, and electromigration
on crystal surfaces, with potentially useful applications
resulting in each case.

We are grateful to Paul Chaikin and Matthew Pelton for 
enlightening discussions.
This research was supported primarily by the National Science Foundation 
through Grants Number DMR-0304906 and DBI-0233971 and in part by the MRSEC 
and REU programs of the NSF
through Grant Number DMR-0213745.


\begin{thebibliography}{17}
\bibitem[{sig()}]{sigma}
\bibinfo{note}{This form applies in both Rayleigh ($a \ll w_0$) and ray-optics
  ($a \gg w_0$) limits \protect\cite{neto00}, and is consistent with our
  measurements, assuming $w_0 \approx \lambda/2$.}
\end{thebibliography}

\end{document}